\documentclass{pasj00}

\begin{document}
\SetRunningHead{Y. Toba \etal\ }{Hyper-luminous Dust Obscured Galaxies discovered by the Hyper Suprime-Cam on Subaru and WISE}
\Received{2014/12/31}
\Accepted{2015/05/25}

\title{Hyper-luminous Dust Obscured Galaxies discovered by the Hyper Suprime-Cam on Subaru and WISE\thanks{Based on data collected at the Subaru Telescope, which is operated by the National Astronomical Observatory of Japan (NAOJ).}}

%
 \author{
   Yoshiki		\textsc{Toba}		\altaffilmark{1},
   Tohru 		\textsc{Nagao}		\altaffilmark{1},
   Michael A.	\textsc{Strauss}	\altaffilmark{2},
   Kentaro		\textsc{Aoki}		\altaffilmark{3}, 
   Tomotsugu 	\textsc{Goto}		\altaffilmark{4},   
   Masatoshi 	\textsc{Imanishi}	\altaffilmark{3,5,6},    
   Toshihiro	\textsc{Kawaguchi}	\altaffilmark{7}, 
   Yuichi 		\textsc{Terashima}	\altaffilmark{8},   
   Yoshihiro 	\textsc{Ueda}		\altaffilmark{9},
   James 		\textsc{Bosch} 		\altaffilmark{2}, 
   Kevin 		\textsc{Bundy}		\altaffilmark{10},    
   Yoshiyuki	\textsc{Doi} 		\altaffilmark{3},   
   Hanae 		\textsc{Inami}		\altaffilmark{11},
   Yutaka 		\textsc{Komiyama}	\altaffilmark{5,6},
   Robert H.	\textsc{Lupton}		\altaffilmark{2},
   Hideo 		\textsc{Matsuhara}	\altaffilmark{12,13},  
   Yoshiki 		\textsc{Matsuoka}	\altaffilmark{5},
   Satoshi 		\textsc{Miyazaki} 	\altaffilmark{5,6},
   Tomoki 		\textsc{Morokuma}	\altaffilmark{14},  
   Fumiaki 		\textsc{Nakata} 	\altaffilmark{3},      
   Nagisa	 	\textsc{Oi}			\altaffilmark{12}, 
   Masafusa		\textsc{Onoue}		\altaffilmark{5,6},
   Shinki 		\textsc{Oyabu}		\altaffilmark{15},
   Paul 		\textsc{Price}		\altaffilmark{2},
   Philip J.	\textsc{Tait} 		\altaffilmark{3},   
   Tadafumi		\textsc{Takata} 	\altaffilmark{5,6},   		   
   Manobu M.  	\textsc{Tanaka} 	\altaffilmark{16},   	      
   Tsuyoshi		\textsc{Terai} 		\altaffilmark{3},      
   Edwin L. 	\textsc{Turner} 	\altaffilmark{2,10},    
   Tomohisa 	\textsc{Uchida} 	\altaffilmark{16},  
   Tomonori 	\textsc{Usuda} 		\altaffilmark{5,6},          
   Yousuke 		\textsc{Utsumi} 	\altaffilmark{17}, 
   Shiang-Yu 	\textsc{Wang} 		\altaffilmark{18},
   Yoshihiko 	\textsc{Yamada} 	\altaffilmark{5}
   }
         
 \altaffiltext{1}{Research Center for Space and Cosmic Evolution, Ehime University, Bunkyo-cho, Matsuyama, Ehime 790-8577}
  \email{toba@cosmos.phys.sci.ehime-u.ac.jp}
  \altaffiltext{2}{Princeton University Observatory, Peyton Hall, Princeton, NJ 08544, USA} 
 \altaffiltext{3}{Subaru Telescope, 650 North A'ohoku Place, Hilo, HI 96720, USA}
 \altaffiltext{4}{Institute of Astronomy and Department of Physics, National Tsing Hua University, No. 101, Section 2, Kuang-Fu Road, Hsinchu 30013, Taiwan}  
 \altaffiltext{5}{National Astronomical Observatory of Japan, Osawa, Mitaka, Tokyo 181-8588} 
 \altaffiltext{6}{Department of Astronomy, School of Science, SOKENDAI (The Graduate University for Advanced Studies), Mitaka, Tokyo 181-8588}
 \altaffiltext{7}{Division of Physics, Sapporo Medical University, S1 W17, Chuo-ku, Sapporo 060-8556}
 \altaffiltext{8}{Department of Physics, Ehime University, 2-5, Bunkyo-cho, Matsuyama, Ehime 790-8577}
 \altaffiltext{9}{Department of Astronomy, Graduate School of Science, Kyoto University, Kitashirakawa-Oiwake cho, Kyoto 606-8502} 
 \altaffiltext{10}{Kavli Institute for the Physics and Mathematics of the Universe  (Kavli IPMU, WPI), Todai Institutes for Advanced Study, the University of Tokyo, Kashiwa 277-8583} 
 \altaffiltext{11}{National Optical Astronomy Observatory, 950 North Cherry Avenue, Tucson, AZ 85719, USA
}
 \altaffiltext{12}{Institute of Space and Astronautical Science, Japan Aerospace Exploration Agency, 3-1-1 Yoshinodai, Chuo-ku, Sagamihara, Kanagawa 252-5210}
 \altaffiltext{13}{Department of Space and Astronautical Science, SOKENDAI (The Graduate University for Advanced Studies), 3-1-1 Yoshinodai, Chuo-ku, Sagamihara, Kanagawa 252-5210}
  \altaffiltext{14}{Institute of Astronomy, Graduate School of Science, The University of Tokyo, 2-21-1 Osawa, Mitaka, Tokyo 181-0015}
 \altaffiltext{15}{Graduate School of Science, Nagoya University, Furo-cho, Chikusa-ku, Nagoya, Aichi 464-8602}
 \altaffiltext{16}{High Energy Accelerator Research Organization (KEK), Institute of Particle and Nuclear Studies, 1-1 Oho, Tsukuba 305-0801} 
 \altaffiltext{17}{Hiroshima Astrophysical Science Center, Hiroshima University, Kagamiyama, Higashi-Hiroshima, Hiroshima, 739-8526}
 \altaffiltext{18}{Institute of Astronomy and Astrophysics, Academia Sinica, No.1, Sec. 4, Roosevelt Rd, Taipei 10617, Taiwan}     


\maketitle

\begin{abstract}
We present the photometric properties of a sample of infrared (IR) bright dust obscured galaxies (DOGs).
Combining wide and deep optical images obtained with the Hyper Suprime-Cam (HSC) on the Subaru Telescope and all-sky mid-IR (MIR) images taken with Wide-Field Infrared Survey Explorer (WISE), we discovered 48 DOGs with $i - K_\mathrm{s} > 1.2$ and $i - [22] > 7.0$, where $i$, $K_\mathrm{s}$, and [22] represent AB magnitude in the $i$-band, $K_\mathrm{s}$-band, and 22 $\micron$, respectively, in the GAMA 14hr field ($\sim$ 9 deg$^2$).
Among these objects, 31 ($\sim$ 65 \%) show power-law spectral energy distributions (SEDs) in the near-IR (NIR) and MIR regime, while the remainder show a NIR bump in their SEDs.
Assuming that the redshift distribution for our DOGs sample is Gaussian, with mean and sigma $z$ = 1.99 $\pm$ 0.45, we calculated their total IR luminosity using an empirical relation between 22 $\micron$ luminosity and total IR luminosity.
The average value of the total IR luminosity is (3.5 $\pm$ 1.1) $\times$ $10^{13} \LO$, which classifies them as hyper-luminous infrared galaxies (HyLIRGs).
We also derived the total IR luminosity function (LF) and IR luminosity density (LD) for a flux-limited subsample of 18 DOGs with 22 $\micron$ flux greater than 3.0 mJy and with $i$-band magnitude brighter than 24 AB magnitude.
The derived space density for this subsample is log $\phi$ = -6.59 $\pm$ 0.11 [Mpc$^{-3}$].
The IR LF for DOGs including data obtained from the literature is well fitted by a double-power law.
The derived lower limit for the IR LD for our sample is $\rho_{\mathrm{IR}}$ $\sim$ 3.8 $\times$ 10$^7$ [\LO \,Mpc$^{-3}$] and its contributions to the total IR LD, IR LD of all ultra-luminous infrared galaxies (ULIRGs), and that of all DOGs are $>$ 3 \%, $>$ 9 \%, and $>$ 15 \%, respectively.
\end{abstract}

\section{Introduction}
\label{Intro}
Recent studies have revealed that almost all massive galaxies harbor a supermassive black hole (SMBH) with a mass of $10^{6-10} M_{\odot}$ at their centers. 
Interestingly, their masses are strongly correlated with those of the spheroid component of their host galaxies, suggesting that galaxies and SMBHs coevolve (e.g., \cite{Magorrian,Marconi}).
How did galaxies and SMBHs form and evolve in the 13.7-billion-year history of the universe? 
The mechanism for the co-evolution of galaxies and SMBHs has not been well-constrained observationally,  although it has been the subject of intense theoretical investigation (e.g., \cite{Di Matteo,Hopkins_08}).
This is in part because many previous studies have been based on optically selected samples.
While deep X-ray (e.g., \cite{Page,Alexander_05,Alexander_11,Ueda_14}), radio (e.g., \cite{Park,Smolcic,Bardelli}), and mid-infrared (e.g., \cite{Stern_05,Lacy_13,Lacy_15}) studies can probe the heavily obscured galaxies that harbor actively growing black holes which are difficult to find optically, they do not cover enough sky to find many of the rarest, most luminous objects. 
For a full understanding of the physics of galaxy--SMBH co-evolution, it is crucial to search for actively accreting galaxy--SMBH systems which may be surrounded by heavy dust.  

Here we focus on ``Dust Obscured Galaxies'' (DOGs: \cite{Dey}) as a key population to tackle the mystery of the co-evolution.
DOGs are very faint in the optical, but are bright in the IR.
The original definition of DOGs was $R - [24] > 7.5$ where $R$ and [24] represent AB magnitudes in the $R$-band and 24 $\micron$, respectively \citep{Dey,Fiore}.
Their mid-IR (MIR) flux densities are three orders of magnitude larger than those at optical wavelengths,
which implies dust heating by significant star formation (SF), an active galactic nucleus (AGN), or both, and the bulk of the optical and ultraviolet (UV) emission from them is absorbed by dust.
There are two sub-classes among the DOGs, characterized by their optical-IR spectral energy distributions (SEDs) \citep{Dey}.  
The so-called ``power-law (PL) DOGs'' show a continuous rise in flux density to longer wavelengths, for which IR color-color plots (e.g., \cite{Melbourne}) imply the presence of AGN activity in their nucleus.
In contrast, the so-called ``bump DOGs'' exhibit a rest-frame 1.6 $\micron$ flux excess that is probably due to the stellar photospheres of cooler stars, which implies that the energy source of their huge IR emission is dominated by SF activity.

A possible key process for the co-evolution is a gas-rich galaxy merger, because the gas accumulating onto the nucleus triggers the quasar activity, whose energy then significantly affects the evolution of their host galaxies (e.g., \cite{Hopkins_06}, see also \cite{Sanders}).
In the context of this galaxy merger scenario, the central BHs and their host galaxies are obscured by a large amount of gas and dust during the initial stage of the co-evolution.
Interestingly, the DOG color criterion is particularly sensitive to heavily reddened galaxies at $z \sim 2$ (e.g., \cite{Desai,Yan}).
This redshift is an interesting epoch because the peak of SF and AGN activity in the Universe and the bulk of stellar mass assembly in galaxies occurred around this epoch (e.g., \cite{Le Floc'h,
Richards,Goto,Bouwens,Courteau,Madau}).

A hydrodynamic simulation conducted by \citet{Narayanan} found that major mergers appear as PL DOGs when the accretion rate on the BH peaks.
According to \citet{Dey}, the fraction of PL DOGs increases with 24 $\mu$m brightness; rising to almost 80\% at 24 $\mu$m flux $>$ 1 mJy, although this fraction has an uncertainty of more than 20\%.
Therefore, IR-bright DOGs in particular could constitute a key population for understanding the co-evolution of galaxies and SMBHs.
However, efficient searches for luminous DOGs have been difficult in previous surveys because (i) these objects are optically too faint to detect in optical bands, and (ii) they are spatially rare ((2.82 $\pm$ 0.05)$\times$ 
10$^{-5}$$h_{70}^{-3}$ Mpc$^{-3}$; Dey et al. 2008) because the timescale of the rapid growth of BHs in the growth phase is much shorter than that of the AGN activity.
Therefore, high-sensitivity and wide-area surveys in both optical and MIR are required to search for the most IR-bright DOGs.

In this study, we perform a systematic search for IR-bright DOGs based on wide and deep images in the optical and MIR.
The optical data are obtained with the Hyper Suprime-Cam (HSC: \cite{Miyazaki}), a gigantic mosaic CCD camera with a 1.5 $\deg$ diameter FoV, which is mounted at the prime focus of the Subaru Telescope.
An ambitious HSC legacy survey is on-going, which will cover a wide area ($\sim$ 1,400 deg$^2$) of the sky\footnote{The HSC legacy imaging survey started in March 2014 as a Subaru strategic program (SSP); 300 nights have been allocated for 5 years in total. This HSC--SSP survey consists of three layers: wide, deep, and ultradeep. This study uses wide-layer data obtained in the 2014 spring run, as described in Section \ref{HSC}. A brief summary of the HSC--SSP may be found  at the following web page: http://www.naoj.org/Projects/HSC/surveyplan.html.}.
The expected sensitivity of this multi-band survey ($g$, $r$, $i$, $z$, and $y$) is $r \sim$ 26 in AB magnitude, almost 40 times deeper than that of the Sloan Digital Sky Survey (SDSS: \cite{York}).
The MIR data are obtained from the Wide-field Infrared Survey Explorer (WISE: \cite{Wright}), which was
launched in 2009.
WISE performed an all-sky survey with a high sensitivity in four bands, 3.4, 4.6, 12, and 22 $\mu$m; we will select DOGs using the 22 $\mu$m data.
The 5$\sigma$ detection limit at 22 $\mu$m is better than 6 mJy \citep{Wright}, which is over an order of magnitude fainter than those of previous IR all-sky survey data (IRAS: \cite{Neugebauer,Beichman}, AKARI: \cite{Murakami,Ishihara}).
The combination of sensitivity and large solid angle of these two surveys allow us to detect mamy DOGs.
\begin{figure}
	\begin{center}
	\includegraphics[width=0.5\textwidth]{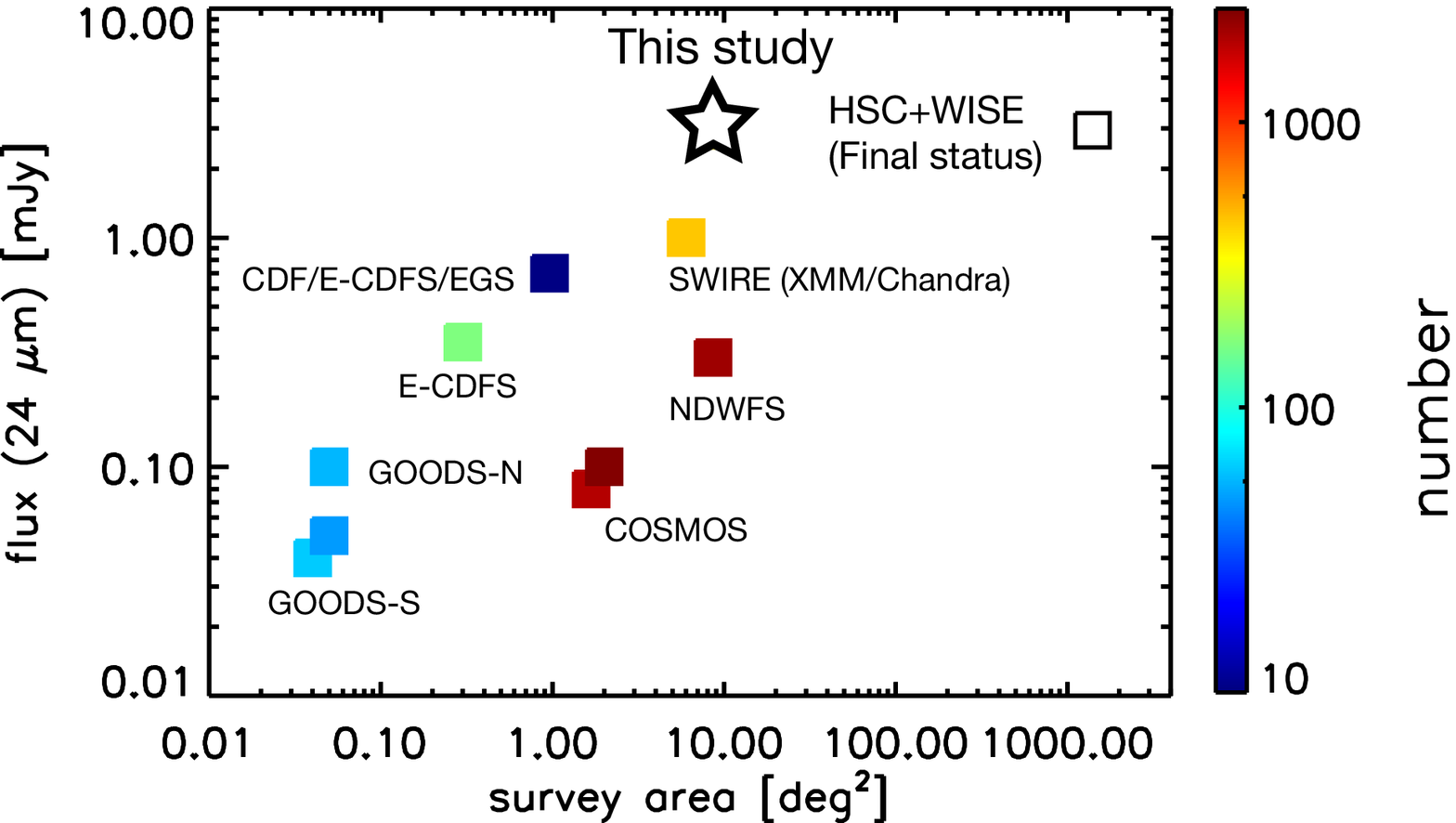} 
	\end{center}
	\caption{Survey area vs. 24 $\micron$ detection limit for previous DOGs surveys (GOODS-S: \cite{Fiore}; GOODS-N: \cite{Pope,Penner}; E-CDFS: \cite{Treister}; CDF/E-CDF/EGS: \cite{Donley}; COSMOS: \cite{Riguccini,Calanog}; SWIRE (XMM/Chandra): \cite{Lanzuisi}; NDFWS: \cite{Dey}). The color bar represents the number of DOGs discovered by each survey. The star represents the survey area and detection limit of this work.}
	\label{survey}
\end{figure}
The survey areas and sensitivities of previous DOGs surveys are summarized in Figure \ref{survey}.
As shown in this figure, the full HSC survey, with WISE, occupies a unique region of parameter space.
In this study, we performed a search for IR-bright DOGs using early HSC data covering $\sim$ 9 deg$^2$, as a benchmark for future HSC studies of DOGs.

It has been suggested that most DOGs represent a subclass of high-redshift ($z \sim$ 2) optically-faint ultra-luminous infrared galaxies (ULIRGs) (e.g., \cite{Melbourne}).
Though ULIRGs are also important objects to study the coevolution of galaxies and SMBHs, systematic searches for high-z ULIRGs are difficult due to the confusion limit of far-IR (FIR) imaging observations.
The situation has been improved with the advent of capable instruments such as the Submillimetre Common-User Bolometer Array 2 (SCUBA2: \cite{Holland}) but the confusion limit is still been severe especially for less luminous ULIRGs.
The Atacama Large Millimeter/submillimeter Array (ALMA) will overcome the confusion problem even for less luminous ULIRGs, although its small field of view means that it cannot carry out wide-angle surveys.
On the other hand, the rarity of DOGs on the sky means that they are not affected by confusion, and thus they are a useful way to study the dusty population at $z >$ 2 statistically.

This paper is organized as follows. Section \ref{Sample_Selection} describes the sample selection of IR-bright DOGs, and their basic properties such as their SEDs are presented in Section \ref{Results}.
In Section \ref{Discussion}, we derive their total IR luminosity function (LF) and estimate their IR luminosity density (LD).
 The cosmology adopted in this paper assumes a flat universe with $H_0$ = 70 km s$^{-1}$ Mpc$^{-1}$, $\Omega_M$=0.3, and $\Omega_{\Lambda}$= 0.7.

\section{Sample Selection}
\label{Sample_Selection}
\begin{figure*}
	\begin{center}
	\includegraphics[width=\textwidth]{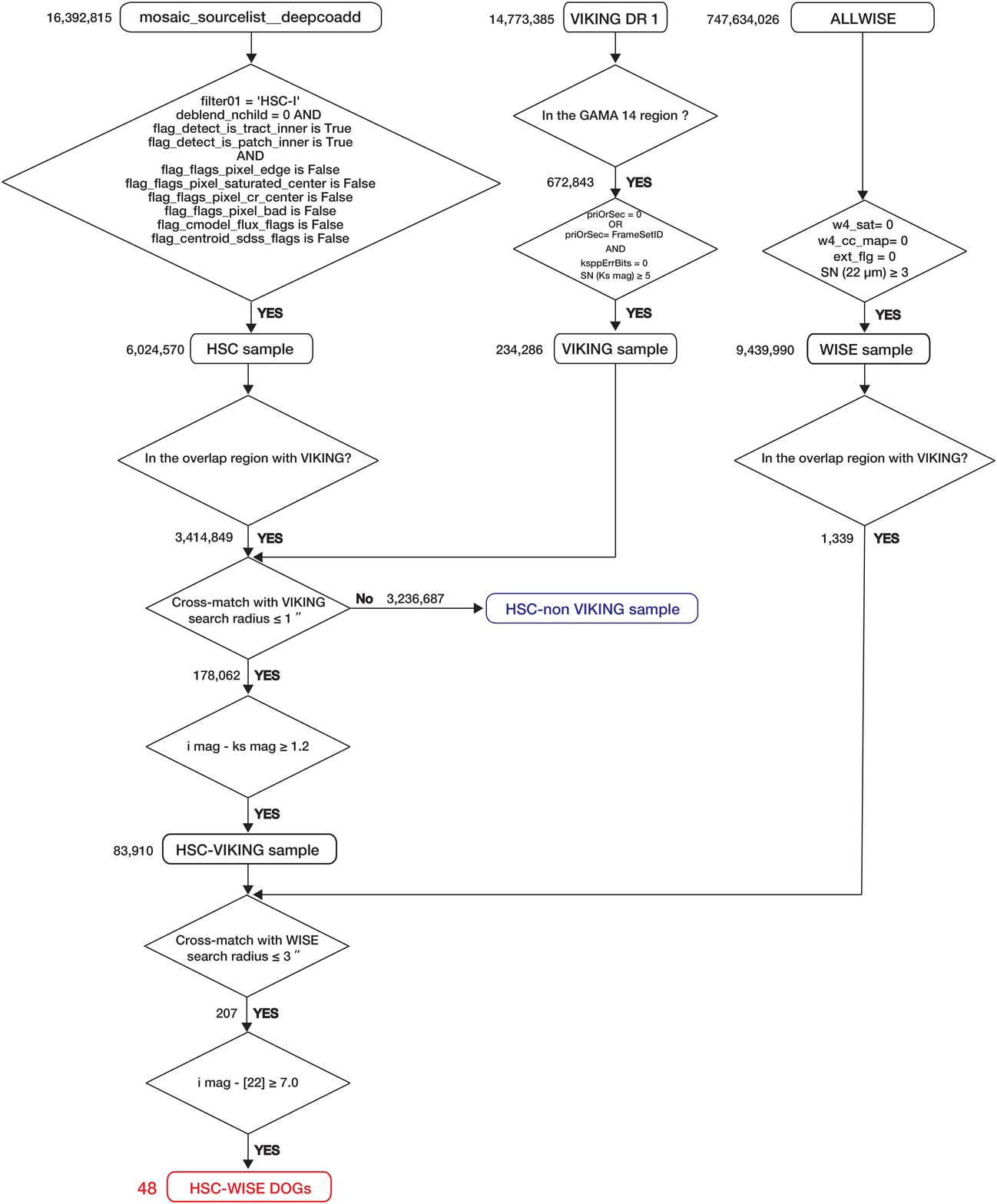} 
	\end{center}
	\caption{Flow chart of the sample selection process.}
	\label{sample_selection}
\end{figure*}

The sample used for this study was selected from WISE MIR sources with optical counterparts detected by  HSC.
A flow chart of our sample selection process is shown in Figure \ref{sample_selection}; we found a total of 48 DOGs.\footnote{For the selection process, we employed the TOPCAT based on the Starlink Tables Infrastructure Library (STIL), which is an interactive graphical viewer and editor for tabular data \citep{Taylor}.}

\begin{figure}
	\begin{center}
	\includegraphics[width=0.43\textwidth]{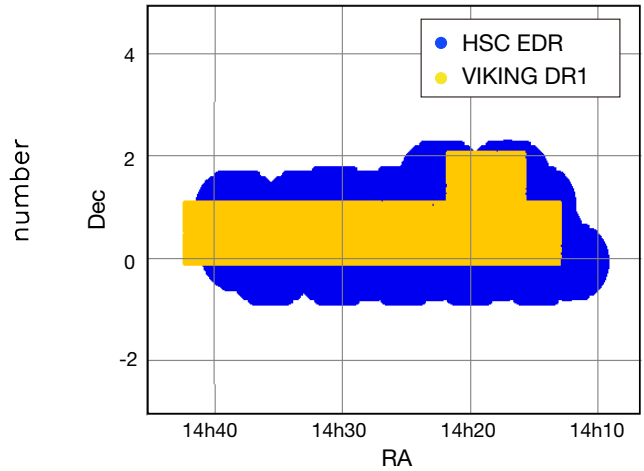} 
	\end{center}
	\caption{Survey footprint for HSC (blue) and VIKING (yellow) data. In this study, we used HSC data in the overlap region with VIKING ($\sim$ 9 deg$^2$).}
	\label{footprint}
\end{figure}

\subsection{HSC sample}
\label{HSC}
In this study, we utilized the HSC--SSP S14A\_0 early data, containing the positions and photometric information of objects detected by observations in March and April in 2014.
The HSC early data were mainly comprised of two catalogs; a wide layer catalog that includes two band photometry ($i$ and $y$), and an ultradeep layer catalog that includes five band photometry ($g$, $r$, $i$, $z$, and $y$).
In particular, we used a sample of 16,392,815 objects detected either in $i$-band or $y$-band in the wide layer catalog which covers the Galaxy And Mass Assembly (GAMA: Driver et al. 2009, 2011) 14hr field (G14: (RA, Dec) $\sim$ (\timeform{216D}, \timeform{0.68D}) with a total survey area of $\sim$ 18.6 deg$^2$, see figure \ref{footprint}).
The expected limiting magnitude (5$\sigma$, \timeform{2''} diameter aperture) for $i$- and $y$- band is approximately 26 and 24 in AB magnitude, respectively. 
The typical full width at half maximum (FWHM) of the point spread function (PSF) for the $i$- and $y$- band data in this field is $\sim$ 0.55 and 0.68 arcsec, respectively.
The data observed by the HSC in the S14A run were analyzed through an early version of the HSC pipeline  (version 2.12.4d\_hsc) using codes from the Large Synoptic Survey Telescope (LSST) software pipeline \citep{Ivezic,Axelrod}, developed by the HSC software team.
This pipeline performs CCD-by-CCD reduction and calibration for astrometry and photometric zeropoints, mosaic-stacking which combines reduced CCD images into a large high S/N coadd image, 
and catalog generation for detecting and measuring sources on the coadd.
The photometric calibration is based on data obtained from the Panoramic Survey Telescope and Rapid Response System (Pan-STARRS) 1 imaging survey \citep{Schlafly,Tonry,Magnier}.
In this study, we employed the cmodel magnitude to estimate $i$-band flux, which is a weighted combination of exponential and de Vaucouleurs fits to the light profile of each object (see \cite{Lupton,Abazajian}).
We limited ourselves to those objects with clean $i$-band photometry and removed duplicates.
In particular, we required:
(i) objects are selected in $i$-band ({\tt filter01} = ``HSC-I''),
(ii) they are completely isolated or were cleanly deblended ({\tt deblend\_nchild} = 0),
(iii) they are unique objects; repeat observations are removed ({\tt flag\_detect\_is\_tract\_inner} = ``True'' and {\tt flag\_detect\_is\_patch\_inner} = ``True''),
(iv) none of the pixels in their footprint are interpolated ({\tt flag\_flags\_pixel\_edge} = ``False''), 
(v) none of the central 3$\times$3 pixels are saturated ({\tt flag\_flags\_pixel\_saturated\_center} = ``False''),
(vi) none of the central 3$\times$3 pixels are affected by cosmic-rays ({\tt flag\_flags\_pixel\_cr\_center} = ``False''),
(vii) there are no bad pixels in their footprint ({\tt flag\_flags\_pixel\_bad} = ``False''),
(viii) there are no problems in measuring cmodel fluxes ({\tt flag\_cmodel\_flux\_flags} = ``False''), and 
 (ix) they have a clean measurement of the centroid ({\tt flag\_centroid\_sdss\_flags} = ``False'' ).
This yields a sample of 6,024,570 $i$-selected sources over $\sim$ 18 deg$^2$.
The $i$-band cmodel AB magnitude ranges from 16.0 to 27.1.

\subsection{WISE sample}
WISE performed an all-sky survey at 3.4, 4.6, 12, and 22 $\micron$, with the PSF FWHM of 6.1, 6.4, 6.5, and 12.0 arcsec.
In this study, we used the latest all-sky ALLWISE catalog \citep{Cutri} that achieved significantly better sensitivity in 3.4 and 4.6 $\micron$ than the WISE all-sky data release \citep{Wright} due to improved data processing.
The 5$\sigma$ photometric sensitivity for 3.4, 4.6, 12, and 22 $\micron$ is better than 0.054, 0.071, 1, and 6 mJy (corresponding to 19.6, 19.3, 16.4, and 14.5 AB magnitudes), respectively.
Note that the number of 12 and 22 $\micron$--detected sources was smaller than earlier release; the improved estimates of the the local background reduced the number of faint objects.

We first created the 22 $\micron$-selected catalog, meaning that we extracted $>$3$\sigma$-detected objects at 22 $\micron$.
We then eliminated sources flagged by the saturation flag, extend flag, and image artifact flag (e.g., diffraction spikes, scattered-light halos, or optical ghosts), leaving a sample with clean photometry.
In this study, we employed the profile-fit magnitude for each band, which ensures reliable photometry for point-sources.
The WISE catalog contains the Vega magnitude of each source, and we converted these to AB magnitude, using offset values $\Delta$m (m$_{\mathrm{AB}}$ = m$_{\mathrm{Vega}}$ + $\Delta$m) for 3.4, 4.6, 12, and 22 $\micron$ of 2.699, 3.339, 5.174, and 6.620, respectively, according to the Explanatory Supplement to the AllWISE Data Release Products \footnote{http://wise2.ipac.caltech.edu/docs/release/allwise/expsup/index.html}.

\subsection{The VISTA Kilo-degree Infrared Galaxy survey (VIKING) sample}
\label{VIKING}
The VISTA Kilo-degree Infrared Galaxy survey (VIKING: \cite{Arnaboldi}) is performing a wide area ($\sim$ 1500 deg$^2$) near-IR (NIR) imaging survey with five broadband filters ($Z$, $Y$, $J$, $H$, and $K_\mathrm{s}$) using the VISTA InfraRed Camera (VIRCAM: \cite{Dalton}) on the VISTA telescope operated by ESO.
The 5$\sigma$ AB photometric sensitivity at $Z$, $Y$, $J$, $H$, and $K_\mathrm{s}$ in \timeform{2''} aperture is 23.1, 22.3, 22.1, 21.5, and 21.2, respectively.
VIKING aims for a seeing FWHM of $<$ 1.0 arcsec \citep{Andrews}.
We used here the data release 1 (DR1), consisting of a total of 14,773,385 sources in 226 deg$^2$.
These data partially overlap with our HSC data.
The overlap between the two covers $\sim$ 9 deg$^2$, as shown in Figure \ref{footprint}; we refer to it as G14--VIKING in what follows.
In this study, we employed the default point source aperture-corrected magnitude (2.0 arcsec diameter) for each band.
We first narrowed the NIR sample to the 672,843 VIKING sources within the G14--VIKING region.
We then extracted the 234,286 sources with signal to noise ratio (S/N) greater than 5 and with clean photometric flags.

\subsection{Cross-identification of HSC and WISE data}
\label{X-id}
\begin{figure*}
	\begin{center}
	\includegraphics[width=0.86\textwidth]{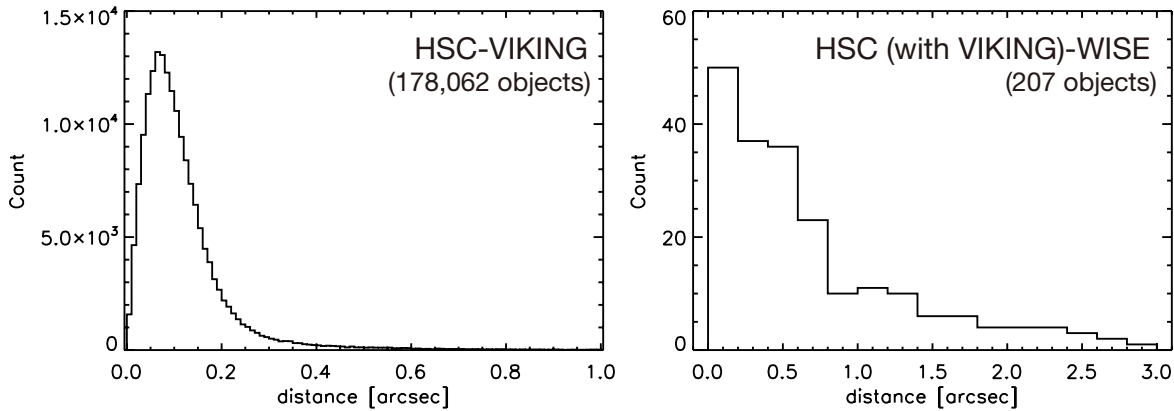} 
	\end{center}
	\caption{Histograms of the angular separation of HSC sources from the VIKING (left) and WISE (right) coordinates. A search radius of 1 arcsec when cross-matching with VIKING and 3 arcsec for WISE was adopted. Cross-matching with the VIKING coordinates selected 178,062 objects within the search radius, while that with the WISE coordinates selected 207 objects.}
	\label{dist}
\end{figure*}
We then cross-identified the HSC sample with the WISE sample.
However, the angular resolutions of HSC and WISE are significantly different, which can give rise to false detections.
For example, a WISE source can have several HSC counterparts candidates lying within the search radius.
We use the fact that DOGs have very red optical-NIR colors.
Hence we first joined the HSC data with NIR data obtained from the VIKING catalog whose angular resolution is roughly comparable to that of the HSC ($<$ 1 arcsec), and we then adopted an optical-NIR color cut to reject sources unlikely to be DOGs before cross-matching with WISE.

\begin{figure}
	\begin{center}
	\includegraphics[width=0.43\textwidth]{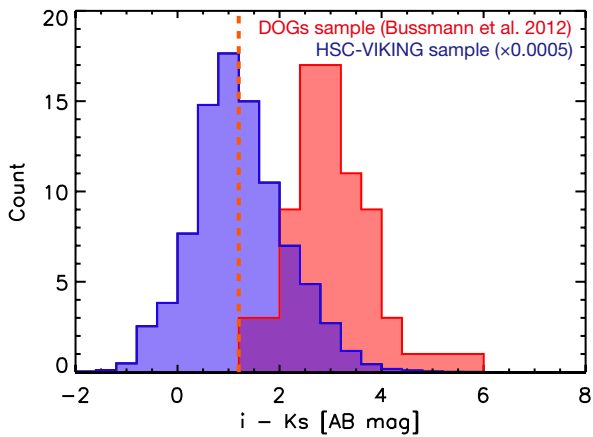} 
	\end{center}
	\caption{$i - K_\mathrm{s}$ color distribution of the sources. The blue histogram represents our matched HSC--VIKING sample (scaled down by a factor of 2000). The red histogram represents the DOGs sample (90 objects) presented by \citet{Bussmann}. The dotted line ($i - K_\mathrm{s} > 1.2$) indicates the adopted threshold for selecting DOGs candidates.}
	\label{i_k}
\end{figure}

First, we narrowed our HSC and WISE samples to sources within the G14--VIKING region, which yields a sample of 3,414,849 HSC sources and of 1,339 WISE sources, respectively.
We then cross-matched the HSC samples with the VIKING sample.
Using a matching radius of 1 arcsec, 178,062 objects were cross-identified, as shown in Figure \ref{dist}; we expect $<$ 0.5 \% of our sources are in correctly matched, given the VIKING surface density.
For this ``HSC--VIKING sample'', we performed a color selection based on $i - K_\mathrm{s}$ color.
\citet{Bussmann} investigated the SEDs of DOGs with spectroscopic redshift in the 8.6 deg$^2$ NDWFS Bo$\mathrm{\ddot{o}}$tes field, which has the requisite multi-band photometry.
Figure \ref{i_k} shows the $i - K_\mathrm{s}$ distribution for 90 DOGs in \citet{Bussmann} and our matched HSC--VIKING sample.
The color distributions of the two are obviously different; the colors of DOGs are much redder than those of other stars and normal galaxies.
In this study, we restrict ourselves to objects with $i - K_\mathrm{s} < 1.2$; 83,910 HSC--VIKING sources met this criterion.

We then cross-matched these 83,910 objects with the WISE sample by using a matching radius of 3 arcsec, which yielded 207 objects as shown in Figure \ref{dist}.
The surface density of WISE sources in the HSC--VIKING sample is approximately 224 deg$^{-2}$, so
adopting this search radius means that the probability of chance coincidence is less than 0.05 \%.
Note that 15 WISE sources have two HSC counterparts because the spatial resolution of the HSC is better than that of WISE. 
In this study, we chose the closest object as the optical counterpart.

\begin{figure}
	\begin{center}
	\includegraphics[width=0.43\textwidth]{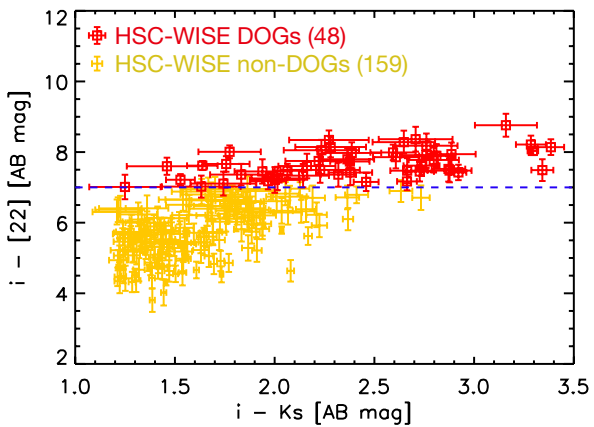} 
	\end{center}
	\caption{Color-color diagram ($i - K_\mathrm{s}$ vs. $i - [22]$) for the HSC--WISE sample. Red squares represent the DOGs that satisfy the color selection criterion ($i - [22] > 7.0$). The remaining sample is represented by yellow symbols. Numbers in () denotes the number of objects.}
	\label{i_k_22}
\end{figure}

For these 207 HSC--WISE objects, we adopted the DOGs selection criterion:
\begin{equation}
i - [22] > 7.0 \;,
\end{equation}
where $i$ and [22] represent AB magnitudes in the HSC $i$-band and WISE 22 $\micron$ band, respectively.
This threshold is consistent with the original DOGs definition, $R$ - [24] $>$ 7.5, as we found by fitting the PL DOGs from \citet{Melbourne} with a single power-law.
Figure \ref{i_k_22} shows the color-color diagram ($i - K_\mathrm{s}$ vs. $i -[22]$) for 207 HSC--WISE objects; 48 DOGs (hereinafter HSC--WISE DOGs) were selected.
\begin{figure}
	\begin{center}
	\includegraphics[width=0.43\textwidth]{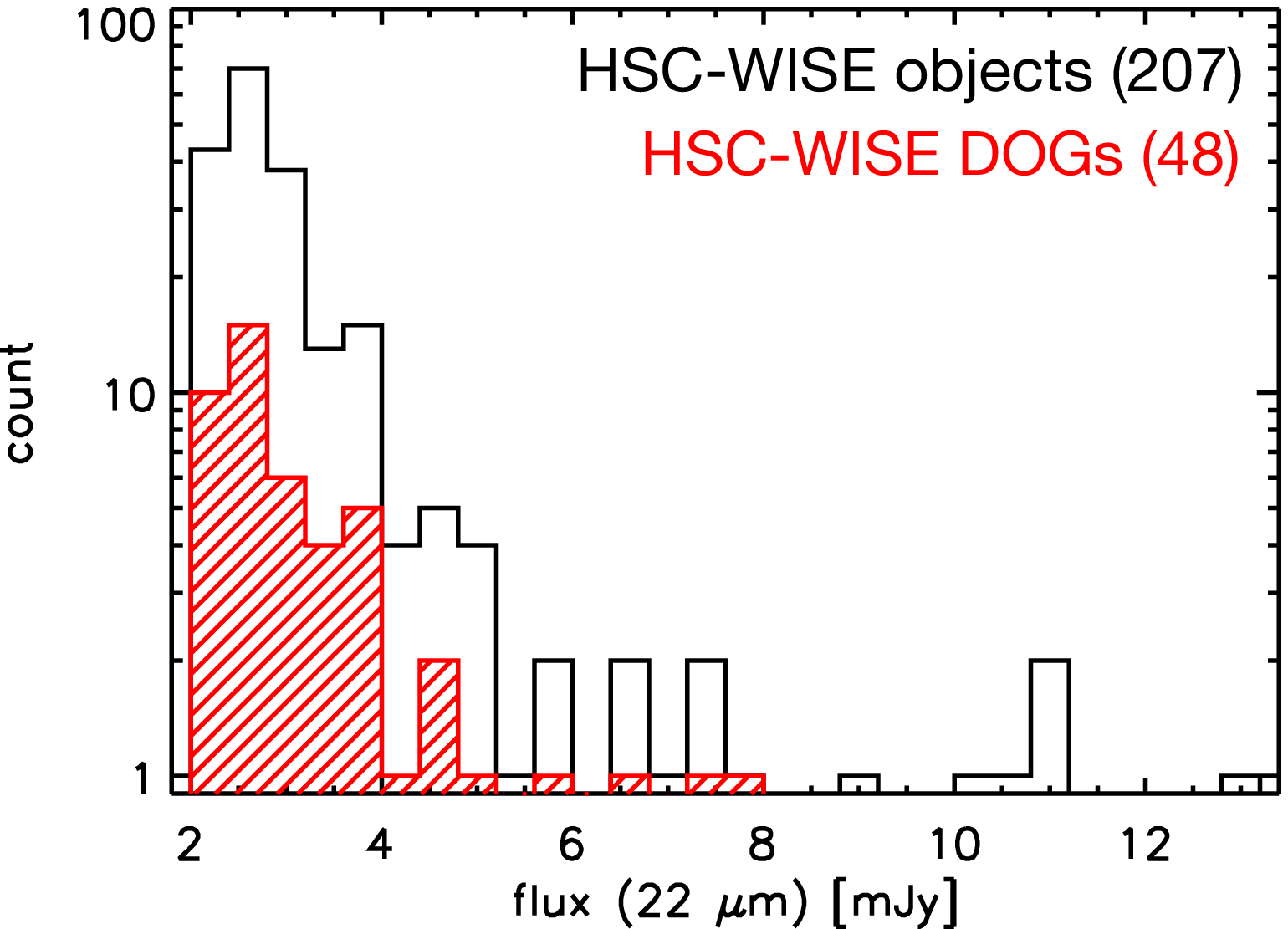} 	
	\end{center}
	\begin{center}
	\includegraphics[width=0.43\textwidth]{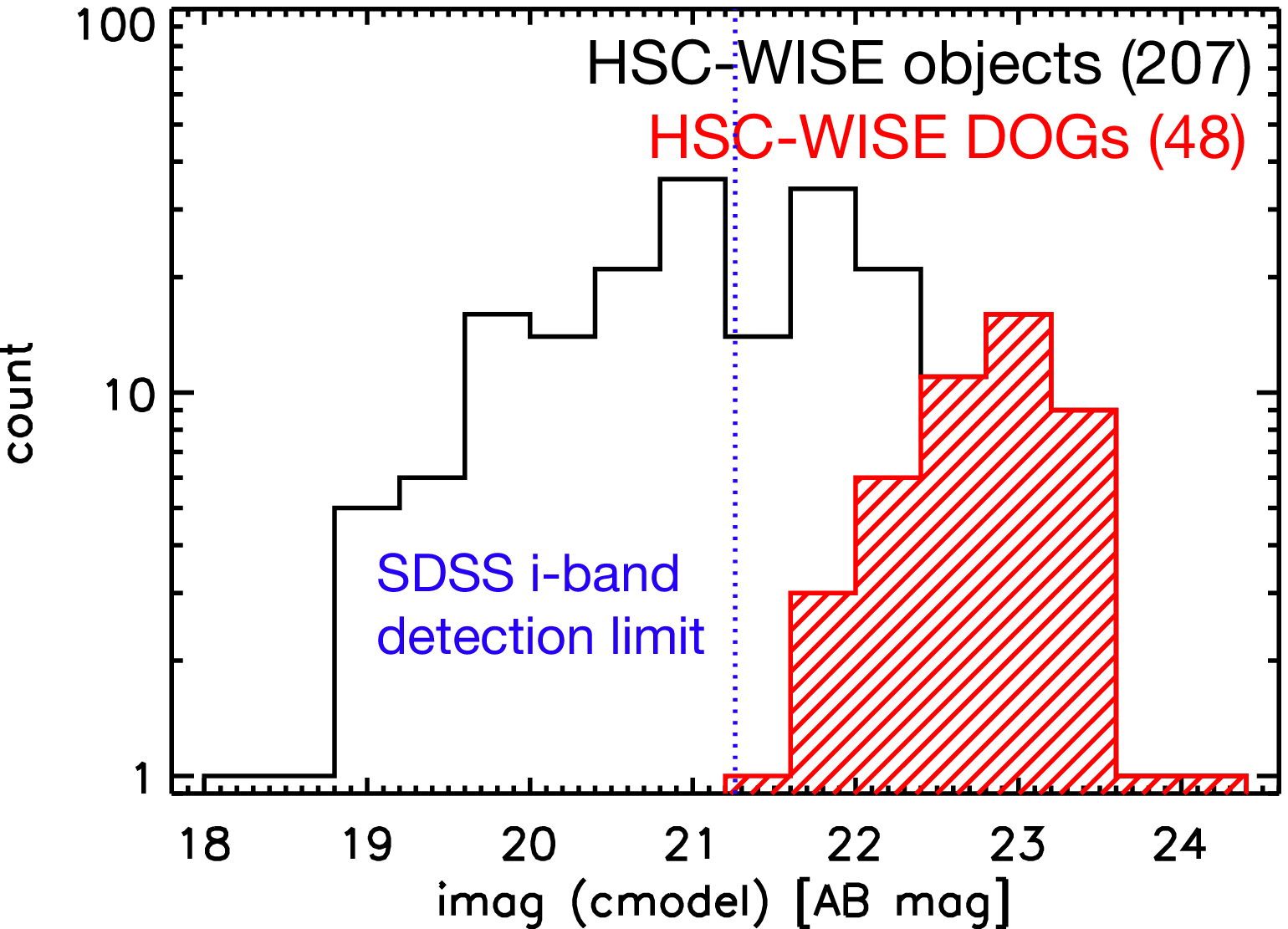} 	
	\end{center}	
	\caption{(top) 22 $\micron$ flux distribution for HSC--WISE objects (black line) and HSC--WISE DOGs (red line). (bottom) the $i$-band magnitude distribution for them. The blue dotted line represents the $i$-band limiting magnitude for the SDSS. Numbers in () denotes the number of objects.}
	\label{fhist}
\end{figure}
Figure \ref{fhist} presents the 22 $\micron$ flux and $i$-band magnitude distributions for our sample. 
Their average and median 22 $\micron$ flux densities are $\sim$ 3.28 and $\sim$ 2.76 mJy, respectively, while their average and median $i$-band magnitude are $\sim$ 22.8 and $\sim$ 22.9, respectively.
We visually inspected the HSC, VIKING, and WISE images of these 48 sources; all are clean detections with good photometry.

If we had not used the prior NIR selection (i.e., $i - K_\mathrm{s} > 1.2$) for HSC objects and simply cross-identified the HSC with WISE data, 310 WISE sources would have more than one HSC counterpart.
We define the ``multiple fraction'', that is, the ratio of the number of WISE sources that have multiple HSC counterparts to that of WISE sources that were cross matched within a search radius of 3 arcsec.
The multiple fraction without and with adopting the prior selection is 310/865 $\sim$ 35.8 \% and 15/207 $\sim$ 7.2 \%, respectively, which indicates our prior selection works well.

We required a VIKING counterpart in our selection, and thus rejected the 3,236,687 HSC sources without a VIKING detection.
We will quantify the incompleteness due to this when we calculate the luminosity function  (Section \ref{reject}).


\section{Results}
\label{Results}
\subsection{Spectral Energy Distributions}
\begin{figure}
	\begin{center}
	\includegraphics[width=0.43\textwidth]{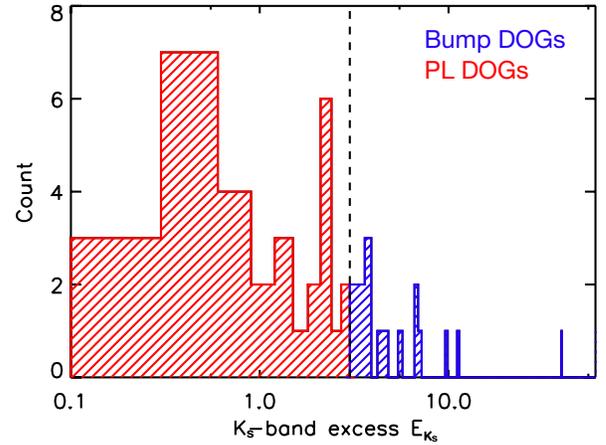} 
	\end{center}
	\caption{The distribution of {\it $K_\mathrm{s}$-band excess} ($E_{K_\mathrm{s}}$) for 48 HSC--WISE DOGs. The threshold to distinguish between PL and Bump DOGs is $E_{K_\mathrm{s}} = 3.0$, which is represented by the dashed line.}
	\label{excess}
\end{figure}
\begin{figure*}
	\begin{center}
	\includegraphics[width=0.8\textwidth]{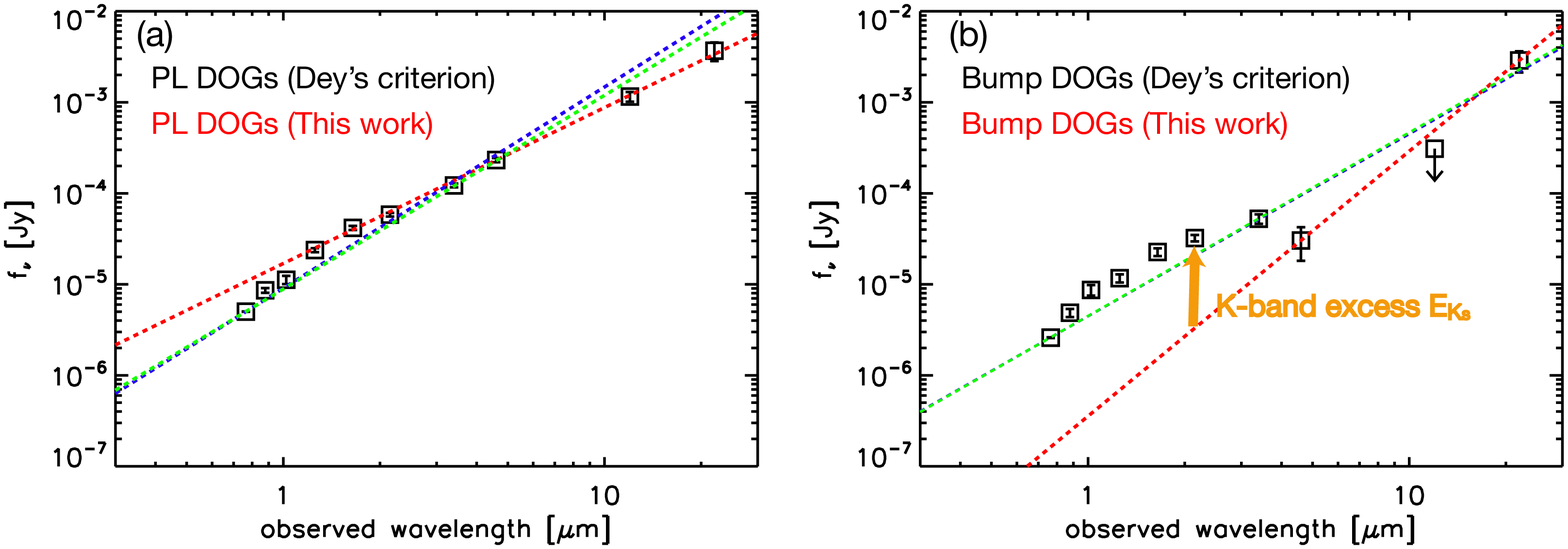} 	
	\end{center}
	\begin{center}
	\includegraphics[width=0.8\textwidth]{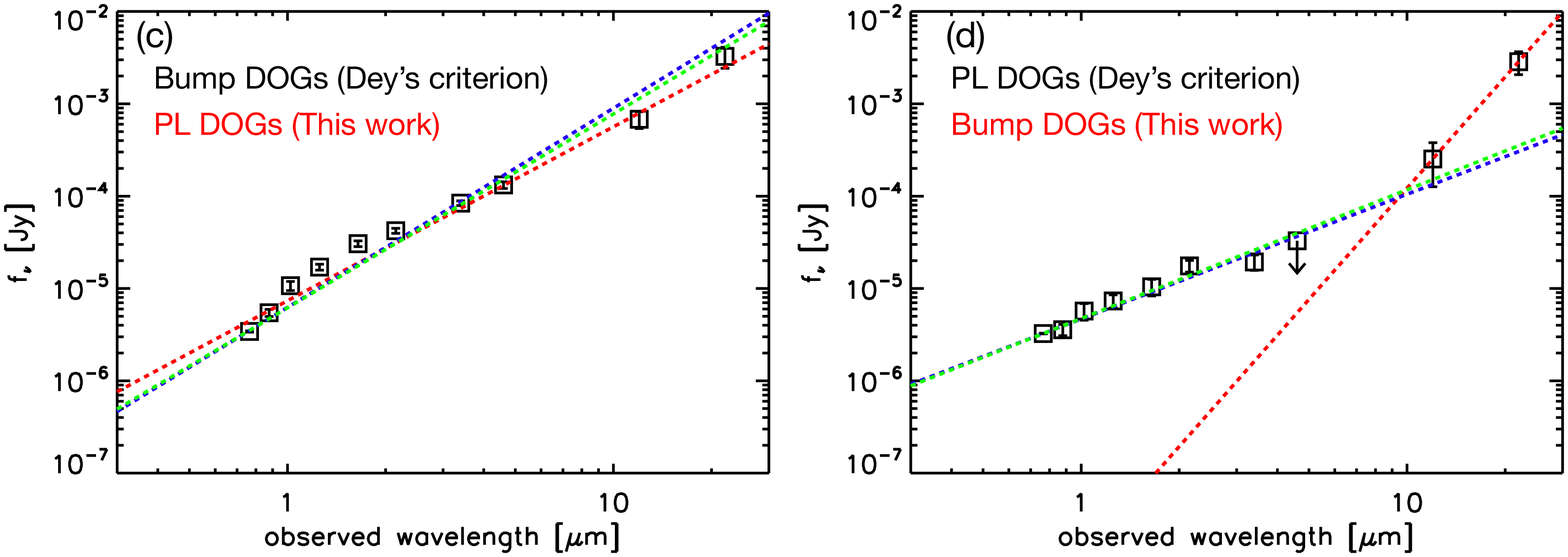}	
	\end{center}	
	\caption{Example SEDs of DOGs in our sample. Red, blue, and green lines represent the best fit linear function determined by fits to longer WISE measurements (4.6, 12, and 22 $\micron$ data), HSC--NIR measurements ($i$, $Z$, $Y$, $J$, $H$, $K_\mathrm{s}$, 3.4, and 4.6 $\micron$ data), and all measurements, respectively. (a,b) Examples of the SED for two objects whose type classification based on the $K_\mathrm{s}$-band excess are consistent with those of \citet{Dey}. (c,d) Examples of the SED for two objects whose type classification are inconsistent with those of \citet{Dey}. Note that the observed SED in (c) fits a power-law well, while that in (d) shows a bump.}
	\label{ex_SED}
\end{figure*}

We classified the 48 DOGs (HSC--WISE DOGs) in our sample into two types (PL and Bump DOGs) based on their observed SEDs.
Note that discriminating the two types is difficult in general because the position of the rest-frame 1.6 $\micron$ bump in their SEDs depends on redshift (see \cite{Melbourne}).
A quantitative classification was originally suggested by \citet{Dey} who performed two power-law fits to the MIR flux measurements of every source, the first to just the Spitzer/IRAC measurements (3.6, 4.5, 5.8, and 8.0 $\micron$ data) and the second to the combined IRAC and MIPS 24 $\micron$ data.
Since this SED discriminant is optimized for the Spitzer data, we define a classification method optimized for our filters ($i$, $Z$, $Y$, $J$, $H$, $K_\mathrm{s}$, 3.4, 4.6, 12, and 22 $\micron$).
In this study, we define the {\it $K_\mathrm{s}$-band excess} to allow us to discriminate PL and Bump DOGs:
\begin{equation}
E_{K_\mathrm{s}} = \frac{f_{K_\mathrm{s}}}{f_{K_\mathrm{s}^{\mathrm{fit}}}},
\end{equation}
where $f_{K_\mathrm{s}}$ is the observed $K_\mathrm{s}$-band flux for each HSC--WISE DOG and $f_{K_\mathrm{s}}^{\mathrm{fit}}$ is the $K_\mathrm{s}$-band flux extrapolated from the power-law fit to the 4.6, 12, and 22 $\micron$ data.
Since our $K_\mathrm{s}$-band data are good quality (S/N $>$ 5, see Section \ref{VIKING}), this value is  a reliable indicator of bump feature.
Figure \ref{excess} shows the $E_{K_\mathrm{s}}$ distribution for HSC--WISE DOGs.
In this study, we classified DOGs with $E_{K_\mathrm{s}}$ greater than 3 as Bump DOGs and all others as PL DOGs, where the threshold was derived empirically by checking each SED. 
Figure \ref{ex_SED} shows examples of their SEDs.
To check our $K_\mathrm{s}$-band excess classification method, we performed two power-law fits to the MIR flux measurements of every source, the first to just the HSC--NIR measurements ($i$, $Z$, $Y$, $J$, $H$, $K_\mathrm{s}$, 3.4, and 4.6 $\micron$ data) and the second to the combined HSC--NIR measurement with WISE 12 and 22 $\micron$ data, in the sprit of \citet{Dey}.
The type classification based on the $K_\mathrm{s}$-band excess is consistent with that based on the two power-law fits for 27 objects (27/48 $\sim$ 56.2 \%) that show a very steep continuum or a clear bump (see (a) and (b) in Figure \ref{ex_SED}).
On the other hand, our classification method is different from that of \citet{Dey} for the remaining 21 objects  (21/48 $\sim$ 43.8 \%), but our method seems to be better from a visual inspection (see (c) and (d) in Figure \ref{ex_SED}). 
Thus our method based on the $K_\mathrm{s}$-band excess works well.

\begin{figure*}
	\begin{center}
	\includegraphics[width=0.8\textwidth]{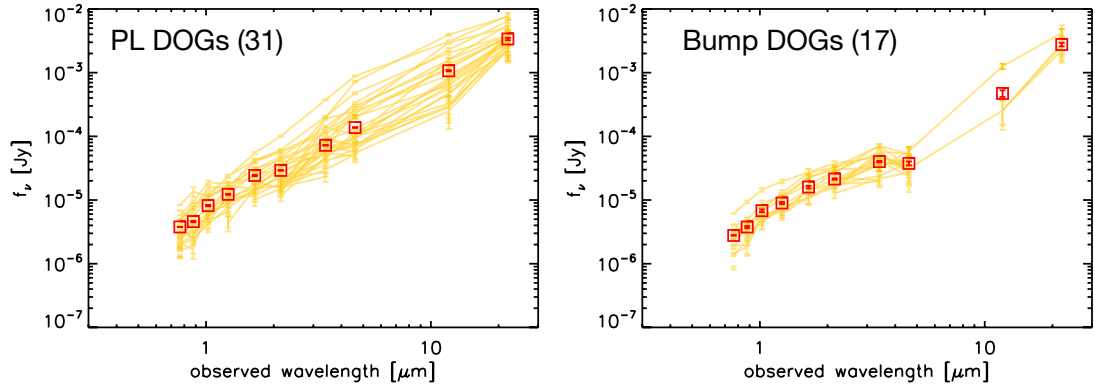} 
	\end{center}
	\caption{The SEDs of PL (left) and Bump (right) DOGs. The red squares represent the average of individual SEDs represented by yellow lines for each type. Numbers in () denotes the number of objects.}
	\label{SED}
\end{figure*}

As a result, 31 ($\sim$ 65 \%) and 17 ($\sim$ 25 \%) objects were classified as PL and Bump DOGs, respectively, which is roughly consistent with those of previous studies (e.g., \cite{Dey}), i.e., IR-bright DOGs tend to be AGN-dominated.
The individual and averaged SEDs for each type are shown in Figure \ref{SED}.

\subsection{DOGs color in WISE}
\label{WISE_color}
\begin{figure*}
	\begin{center}
	\includegraphics[width=0.75\textwidth]{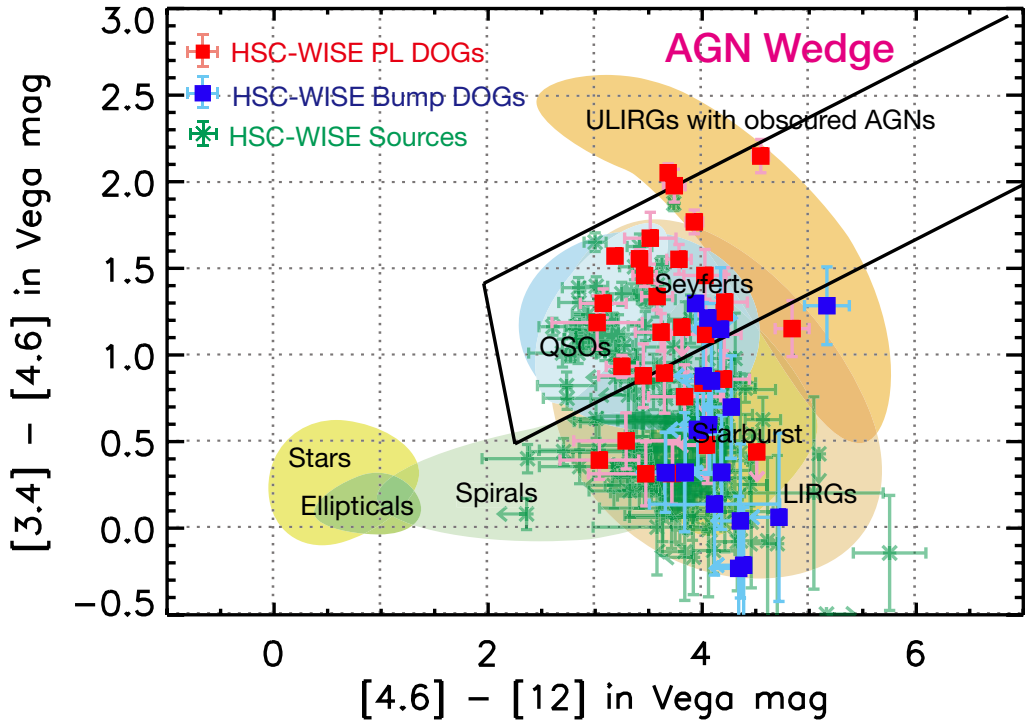} 
	\end{center}
	\caption{WISE color--color diagram of the WISE--HSC objects. Regions with different color shading show typical MIR colors of different populations of objects \citep{Wright}. The solid lines illustrate the AGN selection wedge defined from Mateos et al. (2012, 2013).}
	\label{WISE_color_color}
\end{figure*}

Figure \ref{WISE_color_color} shows the WISE color--color diagram ($[3.4] - [4.6]$ versus $[4.6] - [12]$) for HSC--WISE sources.
The typical MIR colors for various populations of objects are shown with different colors \citep{Wright}.
About 60 \% of PL DOGs are located within the AGN wedge defined by Mateos et al. (2012, 2013), who suggested a highly complete and reliable MIR color selection criteria for luminous AGN candidates based on the WISE and wide-angle Bright Ultrahard XMM-Newton survey (BUXS: \cite{Mateos_12}), while about 73 \% of Bump DOGs are located outside the AGN wedge.
Thus there is not a one-to-one correspondence between AGN classified using Figure \ref{WISE_color_color} and using the PL/Bump criterion, and it is not clear which classification is more reliable.
The WISE colors of AGNs have been well studied  (e.g., \cite{Stern,Assef_13,Toba_14}).
However, our classification is based on broad band SEDs, with up to 10 bands, while the WISE only uses 3 bands.

It is interesting that the HSC--WISE objects are not distributed uniformly in the WISE color--color plane.
The left edge of the distribution of the HSC--WISE objects is due to the $i - K_\mathrm{s}$ selection, thus relatively blue objects such as elliptical and spiral galaxies are removed by the $i - K_\mathrm{s}$ prior criterion.
On the other hand, the steepness of the slope of the continuum in the NIR and MIR regime determines the right edge.
This is because 22 $\micron$-selected objects with extremely steep continuum are difficult to detect at shorter wavelength (3.4, 4.6, and 12 $\micron$), and thus drop out of the sample.
Actually, such extremely red objects have been discovered by WISE.
They are faint or undetected at 3.4 $\micron$ (W1) and 4.6 $\micron$ (W2) while remaining easily detectable at 12 and/or 22 $\micron$.
These objects (so-called ``W12 dropouts'' or ``hot DOGs'': \cite{Wu}) have extremely low surface density ($< 0.03$ deg$^{-2}$), and our HSC--WISE sample does not yet cover enough sky to include such extreme red objects.

\subsection{PL DOGs fraction as a function of 22 $\micron$ flux}
\label{PLfrac_22}
As described in Section \ref{Intro}, some authors have reported that the fraction of DOGs with PL SEDs increases with MIR flux.
If their redshifts are similar, this suggests that the luminous IR sources tend to be more AGN-dominated.
This has been confirmed for ULIRGs by many authors (e.g., \cite{Yuan,Imanishi,Ichikawa}).
\begin{figure}
	\begin{center}
	\includegraphics[width=0.43\textwidth]{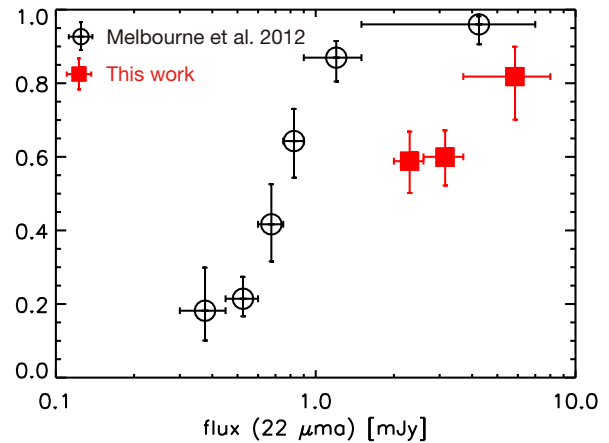} 
	\end{center}
	\caption{The fraction of PL DOGs as a function of 22 $\micron$ flux. The error in the fraction was estimated using binomial statistics (see \cite{Gehrels}). The results of our work (IR-bright DOGs) are represented by filled red squares, while those of DOGs previously discovered by \citet{Melbourne} are represented by open circles.}
	\label{PL_frac}
\end{figure}

The fraction of PL DOGs as a function of 22 $\micron$ flux is shown in Figure \ref{PL_frac} and Table \ref{TB}.
At faint 22 $\micron$ flux, we use the \citet{Melbourne} sample of 113 DOGs using deep Spitzer data in the NOAO Bo$\mathrm{\ddot{o}}$tes field.
We confirmed that the brighter 22 $\micron$ DOGs are more AGN dominated (i.e., PL DOGs), although we found smaller fractions than did \citet{Melbourne}.
Note that the classification in \citet{Melbourne} was performed visually based on the SEDs, while our classification was conducted quantitatively.
Our result based on $K$-band excess could indicate that the PL DOGs fraction for luminous DOGs is not much higher than that expected from previous studies.

\begin{table}
  \caption{Fraction of PL DOGs as a function of 22 $\micron$ flux density.
  \label{TB}}
  \begin{minipage}{0.5\textwidth}
  \begin{center}
    \begin{tabular}{rrr}
      \hline
      22 $\micron$ flux [mJy] & PL DOGs fraction & N\\
      \hline	
      0.375 &   $0.18_{-0.08}^{+0.12}$ \footnotemark[a] & 11\\
      0.525 &   $0.21_{-0.04}^{+0.06}$ \footnotemark[a] & 28\\
      0.675 &   $0.42_{-0.10}^{+0.11}$ \footnotemark[a] & 12\\
      0.825 &   $0.64_{-0.10}^{+0.09}$ \footnotemark[a] & 14\\
	  1.200 &   $0.87_{-0.07}^{+0.04}$ \footnotemark[a] & 23\\       
      2.300 &   $0.59_{-0.09}^{+0.08}$ \footnotemark[b] & 17\\
      3.150 &   $0.60_{-0.19}^{+0.07}$ \footnotemark[b] & 20\\
      4.250 &   $0.96_{-0.08}^{+0.02}$ \footnotemark[a] & 25\\
      5.850 &   $0.86_{-0.05}^{+0.08}$ \footnotemark[b] & 11 \\      
      \hline
      \footnotetext[a]{The fractions were calculated from the data obtained from \citet{Melbourne}.}
      \footnotetext[b]{This work.}      
    \end{tabular}
  \end{center}
  \end{minipage}
\end{table}

\section{Discussion}
\label{Discussion}

\subsection{Total Infrared Luminosity}
We first created a flux-limited subsample of DOGs to determine the luminosity function of these sources.
We extracted the sample with 22 $\micron$ flux greater than 3.0 mJy (below which the WISE 22 $\micron$ data become $<$ 25 \% complete) and $i$-band magnitude smaller than 24.0 in AB mag, which yielded 18 objects.
The 22 $\micron$ flux threshold is responsible for most of the sample reduction; the $i$-band threshold causes only one object to be dropped (see Figure \ref{fhist}).
For $i$-band magnitude, we also determine the threshold by taking the completeness into account.
The completeness of the $i$-band sample is close to 100 \% at the threshold of $i \sim$ 24.

We then estimated the total IR luminosity, $L_{\mathrm{IR}}$ (8--1000 $\micron$) for the 18 DOGs in our flux-limited sample.
We have no spectra of our sources, so we assume that the redshift distribution $P(z)$ is Gaussian, with mean and sigma = 1.99 $\pm$ 0.45 \citep{Dey}.
To estimate the total IR luminosity, we created 1000 simulated 22 $\micron$ luminosity distributions for each source based on $P(z)$.
Note that the MIR flux obtained from deep Spitzer data in \citet{Dey} is much fainter.
So it is not a priori obvious that the redshift distribution should be the same for our sample. 
We checked this using the DOGs sample in \citet{Melbourne}. 
\citet{Melbourne} estimated $L_{\mathrm{IR}}$ based on the broad-band SEDs for 113 DOGs with spectroscopically measured redshifts, including the FIR data of Herschel Space Observatory.
Their sample has a mean spectroscopic redshift of $\sim$ 1.93, similar to \citet{Dey}.
Applying similar flux cuts (i.e., flux (24 $\micron$) $>$ 3.0 mJy and $i$-mag $<$ 24.0) to their sample yields 4 objects, with an average redshift of $\sim$ 1.79 $\pm$ 0.51, which is consistent with that of \citet{Dey} given the uncertainties.
Thus the data are consistent with a redshift distribution of DOGs roughly independent of flux, and we adopt the \citet{Dey} distribution in what follows.  

\begin{figure}
	\begin{center}
	\includegraphics[width=0.43\textwidth]{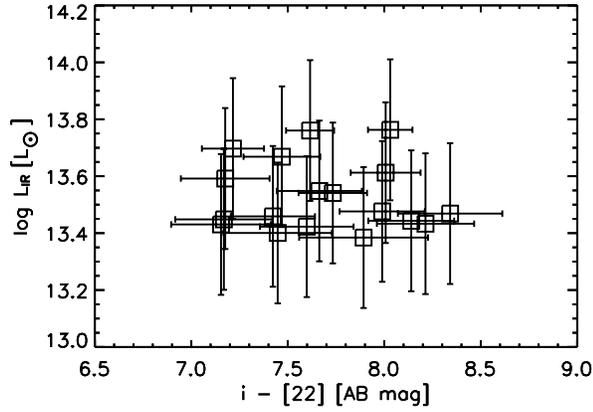} 
	\end{center}
	\caption{Total IR luminosity distribution for 18 DOGs, assuming a redshift distribution ($z$ = 1.99 $\pm$ 0.45) as a function of $i$ - [22]. All objects can be classified as HyLIRGs. The vertical error bars include the redshift uncertainty and the uncertainty in the conversion from MIR luminosity to total IR luminosity.}
	\label{L_IR}
\end{figure}

We also used an empirical relation between observed 24 $\micron$ and total IR luminosity presented by \citet{Melbourne}.
They showed that the $L_{\mathrm{IR}}$ for PL DOGs is well predicted by 24 $\micron$ luminosity with a mean $L_{\mathrm{IR}}$/$\nu L_{\nu}$ (24 $\micron$) = 6.5 $\pm$ 1.4, whereas the scatter for the Bump DOGs was much larger.
We converted from 22 $\micron$ to 24 $\micron$ flux assuming a power-law MIR SED; $f(\nu) \propto \nu^{-\alpha}$, where $\alpha$ was fit to the SED.
We include the uncertainty in this conversion in the $L_{\mathrm{IR}}$ errors.
The derived total infrared luminosity as a function of $i$ - [22] is shown in Figure \ref{L_IR}, where the average value is (3.5 $\pm$ 1.1) $\times$ 10$^{13} \LO$.
This result shows that the IR-bright DOGs discovered by HSC and WISE are hyper-luminous infrared galaxies (HyLIRGs).
Hereinafter, we label these 18 DOGs as ``IR luminous DOGs''.

\subsection{Effect of rejected objects}
\subsubsection{$i$-band undetected DOGs}
In this study, we used an $i$-band selection of HSC objects, and selected DOGs by adopting $i$ - $K_{\mathrm{s}}$ $>$ 1.2 and $i$ - [22] $>$ 7.0.
However, our color cut would allow DOGs to be identified even if they are undetected in $i$-band.
As shown in Figure \ref{fhist} (b), there are no objects with $i$ $>$ 24.0 although the $i$-band detection limit is $\sim$ 26.0.
This is probably because these extreme red DOGs 
(some of them could be $K_\mathrm{s}$-band undetected DOGs, see Section \ref{reject})
are very rare, the area of our survey is too small to detect them.
Therefore, the influence of $i$-band undetected objects on our results is expected to be small.

\subsubsection{$K_\mathrm{s}$-band undetected DOGs}
\label{reject}
When we cross-matched the HSC sample with the VIKING sample, we rejected 3,236,687 HSC sources that did not lie within the 1 arcsec search radius. 
Could some of these rejected objects (HSC-non VIKING sample) be DOGs?
These objects would be too faint at $K_\mathrm{s}$-band to detect with VIKING, but would satisfy the DOGs criterion ($i - [22] > 7.0$). 
There are 39 such objects with 22 $\micron$ flux greater than 3.0 mJy and $i$-band magnitude brighter than 24.0, among the $K_\mathrm{s}$-band non-detections.
Therefore, we may be missing as much as 68 \% of the DOGs due to our $K_\mathrm{s}$-band limit.
Note that the cross-matching the HSC and WISE catalog without using the $i$ - $K_{\mathrm{s}}$ prior cut can give rise to false detections as described in Section \ref{X-id}.
So, this incompleteness is an upper limit.
Actually, 14/39 ($\sim$ 35.9 \%) of DOGs have more than one counterpart candidate, a multiple fraction, which is consistent with that derived in Section \ref{X-id}.
In any case, one should keep in mind this uncertainty when discussing the statistical properties of DOGs in this study.

\subsection{Total Infrared Luminosity Function}
\label{LIR_LF}
Here, we derive the LF following the 1/$V_{\mathrm{max}}$ method described by \citet{Schmidt}.
The 1/$V_{\mathrm{max}}$ method allows us to compute the LF directly from the data; no parameterization or model assumptions are needed.
The volume density $\phi (L)$ and its uncertainty $\sigma_{\phi (L)}$ are derived using the expressions:

\begin{equation}
\phi(L) = \sum_i^N \frac{1}{V_{\mathrm{max},i}}\;,
\end{equation}

\begin{equation}
\sigma_{\phi(L)} =\sqrt{\sum_i^N \frac{1}{V_{\mathrm{max},i}^2}}\;,
\end{equation}
where $V_{{\rm max},i}$ is the maximum co-moving volume that would be enclosed at the maximum redshift at which the $i$th object could be detected.
The sums are over all galaxies falling into a given luminosity bin.
In the context of the cosmology we adopt, $V_{\mathrm{max}}$ is
\begin{equation}
\begin{array}{l}
V_{\mathrm{max}}(z)  =  
\frac{c}{H_0} \int_{\Omega} \int_{z_{\mathrm{min}}}^{z_{\mathrm{max}}} C(z^{\prime}) \frac{(1+z^{\prime})^2 D_A^2}{\sqrt{\Omega_M (1+z^{\prime})^3 + \Omega_{\Lambda}}} \mathrm{d}z^{\prime} \mathrm{d}\Omega,
\end{array}
\label{Vmax}
\end{equation}
where $D_A$ is the angular distance for a given redshift in our adopted cosmology, $\Omega$ is the solid angle of the HSC--VIKING region (9 deg$^2$ $\sim$ 0.003 str), $z_{\mathrm{min}}$ is the lower limit of the redshift bin considered ($z_{\mathrm{min}}$ = 1.99 - 0.45 = 1.54 in this study), and $z_{\mathrm{max}}$ is the maximum redshift at which the object could be seen given the flux limit of the sample, or 1.99 + 0.45 = 2.44, whichever is smaller.
We calculated $z_{\mathrm{max}}$ numerically, as described in Toba et al. (2013, 2014).
Note that when calculating $V_{\mathrm{max}}$, we should take account the detection limit for each survey (HSC, VIKING, and WISE) because our sample is flux-limited.
However, the detection limit adopted for WISE is much larger than those for HSC and VIKING, given our typical SED, we thus assume here that the volume for each object is determined only by the WISE detection limit (flux at 22 $\micron$ = 3 mJy).

The completeness correction function $C(z)$ in equation (\ref{Vmax}) is determined following \citet{Toba_13}.
First, we formulated the dependence of WISE completeness $C(f)$ for flux $f$ obtained from the Explanatory Supplement to the WISE All-Sky Data Release Products\footnote{http://wise2.ipac.caltech.edu/docs/release/allsky/expsup/sec6\_5.html.}.
Secondly, we convert flux to redshift for each object, where luminosity was treated as constant for each object.
While $C(f)$ depends on the region of the sky (see Explanatory Supplement to the WISE All-Sky Data Release Products), it is almost uniform in the region considered in this study (i.e., G14--VIKING region), and we approximate it to be exactly uniform.

\begin{figure}
	\begin{center}
	\includegraphics[width=0.43\textwidth]{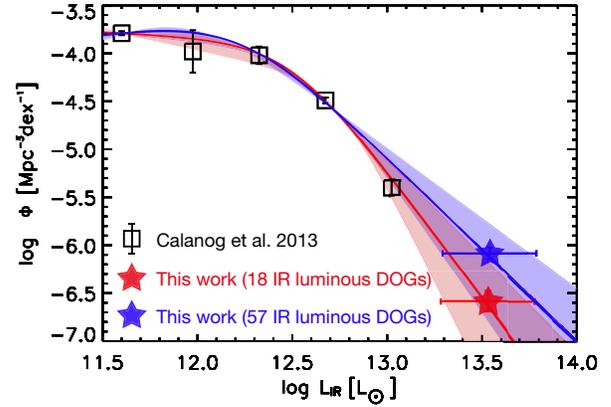} 
	\end{center}
	\caption{The total IR luminosity function for HSC-DOGs at $z \sim$ 2. Data represented by black squares are derived by \citet{Calanog}. Data represented by red and blue stars are our data without and with taking account the 39 DOG candidates without $K_\mathrm{s}$-band detections (see Section \ref{reject}), respectively. The vertical error bars are calculated from the Poisson statistical uncertainty. The red and blue solid lines represent the best fit double-power law function for all data points without and with including the 39 DOGs without $K_\mathrm{s}$-band detections, respectively. The red and blue shaded regions represent the uncertainties of the resultant best fit function due to the large uncertainty in the IR luminosity at bright end.}
	\label{LF}
\end{figure}

The total IR LF (i.e., the volume density of the galaxies per unit luminosity range) of our HSC--WISE DOGs  computed with the 1/$V_{\mathrm{max}}$ method, is shown in Figure \ref{LF}.
The derived space density from the IR LF for our sample (18 DOGs) is log $\phi$ = $-$6.59 $\pm$ 0.11 [Mpc$^{-3}$].
Figure \ref{LF} also shows the LF of DOGs presented by \citet{Calanog} who discovered 3077 DOGs with 24 $\mu$m flux greater than 0.1 mJy in the 2 deg$^2$ of the Cosmic Evolution Survey.
We fit the LF for all galaxies using the double-power law (e.g., \cite{Marshall}):
\begin{equation}
\phi(L)\mathrm{d}L = \phi^* \left\{ \left( \frac{L}{L^*} \right)^{-\alpha} +  \left( \frac{L}{L^*} \right)^{-\beta} \right\}^{-1} \frac{\mathrm{d}L}{L^*}.
\end{equation}
The free parameters are the characteristic luminosity $L^*$, the normalization factor $\phi^*$, 
the bright-end slope $\alpha$, and the faint-end slope $\beta$, respectively. 
The best-fitting values are summarized in Table \ref{best-fitt_LF}, and we found that the shape of the LF can be well fitted by the double-power law form.
In particular, the bright side of the LF is not exponential in shape, predicted e.g., by the \citet{Schechter} function.
This is probably due to the contribution of AGN-dominated DOGs, reflecting the fact that the AGN LF can be fitted by a double-power law (e.g., \cite{Ueda,Richards,Assef}).

As mentioned in Section \ref{reject}, we may be missing as many as 39 DOGs due to our $K_{\mathrm{s}}$-band limit.
We also derived the LF including all 39 DOG candidates.
The resultant fitting values are summarized in Table \ref{best-fitt_LF}.
When including the 39 DOG candidates in our sample (i.e., a total of 57 DOGs), their volume density and bright-end slope are log $\phi$ = $-$6.09 $\pm$ 0.06 [Mpc$^{-3}$] and $\alpha$ = -1.9 $\pm$ 0.1, respectively, meaning that the volume density increases and the bright-end slope gets flatter.
The bright end slope is also sensitive to the uncertainty of the IR luminosity at the bright end.

 \begin{table*}
  \caption{Best double-power law fit parameters for the total IR LF of DOGs.}
  \label{best-fitt_LF}
  \begin{minipage}{\textwidth}
  \begin{center}
    \begin{tabular}{ccccc}
      \hline
	 &  $\phi$ [Mpc$^{-3}$dex$^{-1}$] & L$_{\mathrm{IR}}^*$ [$\LO$] & $\alpha$ (bright-end) & $\beta$ (faint-end) \\
\hline
18 IR luminous DDGs & $(1.3 \pm 0.6) \times 10^{-4}$ & $(3.1 \pm 0.8) \times 10^{12}$ & $-2.6 \pm 0.2$ & $-0.1 \pm 0.2$\\  
     \hline
57 IR luminous DOGs \footnotemark[*] & $(3.1 \pm 1.9) \times 10^{-4}$ & $(1.2 \pm 0.5) \times 10^{12}$ & $ -1.9 \pm 0.1$ & $0.5 \pm 0.7$\\  
    \hline       
    \footnotetext[*]{Including the 39 possible DOGs that could be missed in the sample selection (see Section \ref{reject}).}
    \end{tabular}
  \end{center}
  \end{minipage}
\end{table*}

\subsection{Total Infrared Luminosity Density}
\label{LIR_LD}
\begin{figure*}
	\begin{center}
	\includegraphics[width=0.75\textwidth]{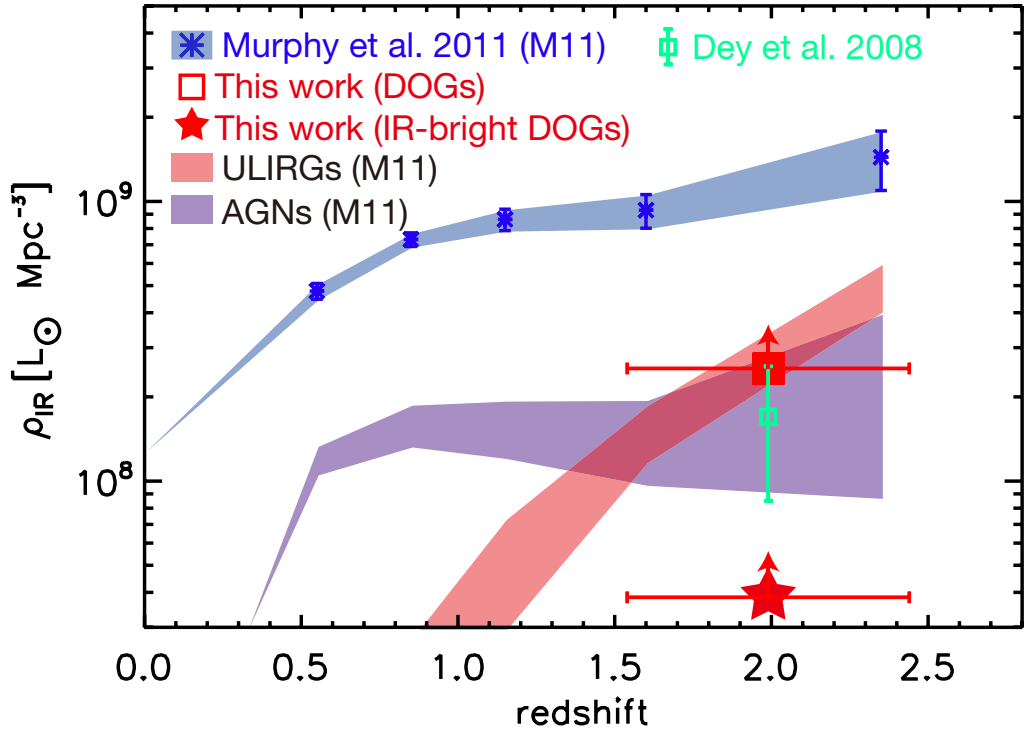} 
	\end{center}
	\caption{The lower limit of infrared luminosity density (IR LD) for HSC-DOGs. Blue points represent total IR LD obtained from \citet{Murphy}. Purple and red shaded regions represent the IR LD contributed by AGNs and ULIRGs, respectively. The IR LD for DOGs of all luminosities including our sample is shown as a red square, and the IR LF for IR luminous DOGs is shown as a red star. The green square shows the IR LD obtained from \citet{Dey}}
	\label{IRLD}
\end{figure*}
One of the primary purposes in computing the IR LF is to estimate the IR luminosity density (LD), which is thought to be a good tracer of SF/AGN activity hidden by dust.
In order to estimate the IR LD contributed by the DOGs, we integrated the IR LF weighted by the luminosity, i.e.,
\begin{equation}
\rho_{\mathrm{IR}} = \int L_{\mathrm{IR}}\phi(L) \mathrm{d}L.
\end{equation}
In this study, we integrated the best-fitting double-power law function over the luminosity range of 10$^{11}$ $<$ $L_{\mathrm{IR}}$ [\LO] $<$ 10$^{14}$.
Given the uncertainties in the IR LF at the bright end and the completeness of our sample, we simply estimated a lower limit to the IR LD of (IR luminous) DOGs.
Figure \ref{IRLD} shows the resultant IR LD ($\rho_{\mathrm{IR}}$) for DOGs.
The value and its contribution to other populations (see below) are summarized in Table \ref{cont}.
For comparison, we also show the total IR LD (i.e., before subtracting out an estimate of the AGN contribution), $\rho_{\mathrm{IR}}$ for AGNs and ULIRGs derived from \citet{Murphy}.
The derived lower limit of IR LD for DOGs, $\rho_{\mathrm{IR}}$ (DOGs), is $\sim$ 2.5 $\times$ 10$^8$ 
 [\LO \,Mpc$^{-3}$], in good agreement with that estimated from \citet{Dey}.
We find that DOGs contribute at least $\sim$ 20 \%  and $\sim$ 61 \% of the total IR LD and ULIRGs IR LD, respectively, which is roughly consistent with the comparison between the DOGs sample of \citet{Dey} and the 24 $\mu$m-selected sample of \citet{Caputi}.
Note that we simply conducted a linear interpolation for data \citep{Murphy} when estimating the total and ULIRGs' IR LD at $z$ = 1.99.

In this study, we also calculated the IR LF for IR luminous DOGs ($L_{\mathrm{IR}}$ $>$ 10$^{13}$ \LO, i.e., the objects in our sample), and thus estimated their contribution to the IR LD of all DOGs and ULIRGs.
The resultant IR luminous DOGs LD of lower limit is $\sim$ 3.8 $\times$ $10^7$  [\LO \,Mpc$^{-3}$] and its contribution to $\rho_{\mathrm{IR}}$ (total), $\rho_{\mathrm{IR}}$ (ULIRGs), and  $\rho_{\mathrm{IR}}$ (DOGs) are $>$ 3 \%, $>$ 9 \%, and $>$ 15 \%, respectively.
The derived IR LD contributes less to the total IR LD at $z \sim 2$ while their contribution to IRLD of DOGs is still moderately large.

\begin{table*}
  \caption{The lower limit of the contribution of the IR LD for DOGs and IR luminous DOGs to other population.
  \label{cont}}
  \begin{center}
    \begin{tabular}{c|c|rrr}
      \hline	
	 &  $\rho_{\mathrm{IR}}$ & \multicolumn{3}{c}{contribution [\%]} \\
     &  [\LO \,Mpc$^{-3}$] & $\rho_{\mathrm{IR}}$ (DOGs) & $\rho_{\mathrm{IR}}$ (ULIRGs) &  $\rho_{\mathrm{IR}}$ (total) \\
      \hline	
	  DOGs & $\sim$ 2.5 $\times$ 10$^8$ & --- & $\sim$ 61 & $\sim$ 20 \\ 	
      \hline
      IR luminous DOGs & $\sim$ 3.8 $\times$ $10^7$ & $\sim$ 15 & $\sim$ 9 & $\sim$ 3\\ 
      \hline
    \end{tabular}
  \end{center}
\end{table*}

\section{Summary}
Using early HSC-wide survey data and the WISE MIR all-sky survey data, we performed a search for IR-bright DOGs in the GAMA 14hr-VIKING region ($\sim$ 9 deg$^2$).
We first created clean subsamples of HSC and WISE data, and also created a NIR selected subsample obtained from VIKING.
We cross-identified HSC with the NIR subsample using a threshold inferred from those of previously discovered DOGs to avoid mis-identification when cross-matching HSC with WISE directly.
We then cross-identified with WISE data and adopted the DOGs color selection ($i - [22] > 7.0)$, which yielded 48 DOGs.
For those DOGs, we investigated their photometric properties as well as their statistical properties by  constructing the IR LF and by estimating their IR LD.
The main results are as follows:
\begin{enumerate}
\item Among 48 DOGs, 31 DOGs ($\sim$ 65 \%) are PL DOGs and 17 DOGs ($\sim$ 25 \%) are bump DOGs, according to their SEDs (Section 3.1). The WISE colors for PL DOGs indicate that they harbor AGNs in their nucleus (Section 3.2).
\item The fraction of PL DOGs increases with increasing 22 $\micron$ luminosity (Section 3.3).
\item Assuming that the redshift distribution for our DOGs sample is Gaussian, with mean and sigma $z$ = 1.99 $\pm$ 0.45, the average total IR luminosity of them is (3.5 $\pm$ 1.1) $\times$ $10^{13} \LO$. Thus these objects are recognized as HyLIRGs (Section 4.1).
\item The derived space density from the IR LF for our sample (18 DOGs) is log $\phi$ = $-$6.59 $\pm$ 0.11 [Mpc$^{-3}$], and the IR LF for DOGs including data obtained from the literature is well fitted by a double-power law with $L_{\mathrm{IR}}^*$ = (3.1 $\pm$ 0.8) $\times 10^{12}$ \LO and bright-end slope $\alpha$ = -2.6 $\pm$ 0.2 (Section 4.3).
\item The derived lower limit of IR LD for our sample is $\sim$ 3.8 $\times$ 10$^7$ [\LO \,Mpc$^{-3}$], and its contributions to the total IR LD, IR LD of all ULIRGs, and that of all DOGs are $>$ 3 \%, $>$ 9 \%, and $>$ 15 \%, respectively (Section 4.4).
\end{enumerate}

The current sample is based on only 9 deg$^2$ of imaging data. 
The HSC survey will cover more than 100 times as much sky, 1400 deg$^2$, when it is complete, allowing the identification of roughly 3300 DOGs.  
This sample will be large enough to carry out a variety of statistical analyses to understand the physical nature of these objects.  
In particular, cross-correlation with the BOSS spectroscopic quasar sample \citep{Paris} will allow a proper determination of the redshift distribution of the sample, and the five-color optical photometry will give independent constraints on the redshift.   
In fact, our objects are bright enough to allow spectroscopy to be performed.
We will also search for DOGs in the deep fields (27 deg$^2$) of the HSC survey: going a magnitude deeper in optical filters, it will allow identification of even redder sources than are explored in this paper.




\bigskip
The authors gratefully acknowledge the anonymous referee for a careful reading of the manuscript and very helpful comments.
We are also deeply thankful to Drs Hisanori Furusawa, Sogo Mineo, and Michitaro Koike (NAOJ) who kindly helped us regarding the treatment of the HSC data.
We also thank Dr Masayuki Akiyama (Tohoku University) for fruitful discussions.
This publication makes use of data products from the Wide-field Infrared Survey Explorer, which is a joint project of the University of California, Los Angeles, and the Jet Propulsion Laboratory/California Institute of Technology, funded by the National Aeronautics and Space Administration. 
This work was achieved by the grant of the Research Assembly supported by the Research Coordination Committee, National Astronomical Observatory of Japan (NAOJ).
This work is also based on data obtained from the ESO Science Archive Facility under request number (toba, \#136654) regarding the VIKING data.
This paper makes use of software developed for the Large Synoptic Survey Telescope. We thank the LSST Project for making their code available as free software at http://dm.lsstcorp.org.
The Pan-STARRS1 Surveys (PS1) have been made possible through contributions of the Institute for Astronomy, the University of Hawaii, the Pan-STARRS Project Office, the Max-Planck Society and its participating institutes, the Max Planck Institute for Astronomy, Heidelberg and the Max Planck Institute for Extraterrestrial Physics, Garching, The Johns Hopkins University, Durham University, the University of Edinburgh, Queen's University Belfast, the Harvard-Smithsonian Center for Astrophysics, the Las Cumbres Observatory Global Telescope Network Incorporated, the National Central University of Taiwan, the Space Telescope Science Institute, the National Aeronautics and Space Administration under Grant No. NNX08AR22G issued through the Planetary Science Division of the NASA Science Mission Directorate, the National Science Foundation under Grant No. AST-1238877, the University of Maryland, and Eotvos Lorand University (ELTE).
T.N. acknowledges financial supports from the Japan Society for the Promotion of Science (grant no. 25707010) and from the Yamada Science Foundation.
This work was supported by World Premier International Research Center Initiative (WPI Initiative), MEXT, Japan.





\end{document}